\documentclass[reprint,aps,amsmath,amssymb,]{revtex4-2} %new
\usepackage{graphicx}% Include figure files %new
\usepackage{dcolumn}% Align table columns on decimal point %new
\usepackage{bm}% bold math %new
\usepackage{xcolor}

\begin{document}

\title{Spatiotemporal control of entangling gates on atomic N-qubit systems}

\author{Ignacio R. Sola}
\email{corresponding author: isolarei@ucm.es}
 \affiliation{Departamento de Quimica Fisica, Universidad Complutense, 28040 Madrid, Spain}%Lines break automatically or can be forced with \\
  % \email{isolarei@ucm.es}
\author{Seokmin Shin}
 \affiliation{School of Chemistry, Seoul National University, 08826 Seoul, Republic of Korea}
\author{Bo Y. Chang}
 \email{corresponding author: boyoung@snu.ac.kr}
 \affiliation{School of Chemistry, Seoul National University, 08826 Seoul, Republic of Korea}
 \affiliation{Research Institute of Basic Sciences, Seoul National University, 08826 Seoul, Republic of Korea}
%   \email{corresponding author: boyoung@snu.ac.kr}

% title: C-PHASE with transformed-limited pulse sequences using structured light
\begin{abstract}
We use a novel optimization procedure that includes the temporal and 
spatial parameters of the pulses acting on arrays of trapped neutral atoms, 
to prepare entangling gates in $N$-qubits systems.
The spatio-temporal control allows treating %to treat 
a denser array of atoms, where
each pulse acts on a subset of the qubits, potentially allowing to speed the
gate operation by two orders of magnitude by boosting the dipole-blockade
between the Rydberg states.
Studying the rate of success of the algorithm under different constraints, 
we evaluate the impact of the proximity of the atoms and, indirectly, the role of 
the geometry of the arrays in $3$ and $4$-qubit systems, 
as well as the minimal energy requirements and how this energy is used among the different qubits.
Finally, we characterize and classify all optimal protocols according to the
mechanism of the gate, using a quantum pathways analysis.
\end{abstract}

\maketitle

\section{Introduction}

Atoms trapped by optical tweezers \cite{Browaeys_PRX2014,Browaeys_Science2016,Wilson_PRL2022,Thompson_PRXQ2022,Ahn_OptE2016}, interacting through Rydberg blockade \cite{Comparat_JOSAB2010,Tong_PRL2004,Saffman_NPhys2009,Grangier_Nphys2009,Adams_PRL2010},
 can be used to generate few multi-particle entanglement \cite{Lukin_PRL2018,Zhan_PRL2017,Ahn_PRL2020,Grangier_PRL2010,Saffman_Nature2022,Saffman_PRA2015,
Saffman_PRA2010,Picken_QCT2018,Malinovsky_PRA2004,Malinovsky_PRL2004,Malinovsky_PRL2006} 
and simple quantum circuits  \cite{Jaksch_PRL2000,Saffman_QIP2011,Saffman_PRA2015,Lukin_PRL2001,Lukin_PRL2019,Cohen_PRXQ2021,
Shi_QSTech2022,Shi_PRApp2018,Adams_PRL2014,Adams_JPB2019,Malinovsky_PRA2014,Goerz_JPB2011,
Morgado_AVSQSci2021,Alexey_PRL2021,Sanders_PRA2020}. 
However, to go further in the quest of the  quantum computer \cite{nielsen_chuang_2010,Saffman_RMP2010}, 
one needs to improve the system addressability and controllability \cite{Browaeys_NP2020}. 

While %the technology
technology has evolved to control the position of the atoms in optical traps
with great precision in arrays of any dimension,
current setups typically %the typical designs 
use ordered arrays of  largely separated atoms, which form independent qubits.
Entangling gates, based on the dipole-blockade mechanism, with energy $d_{\cal B}$, 
%depending on the inverse of the distance between the atoms to the third or six power,
require time durations larger than $\hbar d_{\cal B}^{-1}$, and hence operate in time-scales 
near the microsecond regime for atoms separated over $\gtrsim 5 \mu$m.
%One of the problems in the usual designs with ordered arrays of atoms largely separated in optical traps, is the long duration of the two-qubit gates, that must operate in the micro- or sub-microsecond regime  to avoid exciting more than one atom in a Rydberg state using the dipole blockade mechanism. 
Optimal control theory can be used to find efficient and robust pulse sequences \cite{Shapiro_PRA1993,Kosloff_PRA1994,Rabitz_PRL1992,Rabitz_NJP2010}, but 
in spite of increasing the complexity of the pulse features in the time domain,
it remains very challenging to accelerate the gates without changing the setup 
\cite{Sauvage_PRL2022,Khazali23}.
%\tcr{que significa "without changing the set-up"}
% ADD REFERENCES OF CONTROL APPLICATIONS TO GATES 
We have recently proposed a different approach to deal with this problem 
by controlling not only the time-domain features of the pulses but also
the spatial profiles of the laser beams or the position of the atoms with respect
to the fields, acting over relatively close, but no longer independent, qubits.

%by using denser arrays of that allow to boost the dipole-blockade.  The price to pay is that the qubits can no longer be regarded as independent,  as the laser beams may overlap significantly with more than one qubit site. This can be regarded as a problem or as an opportunity, since by controlling the position of the atoms with respect to the different laser beams, adding a spatial control knob into the problem, one gains a novel and important feature  that may provide both flexibility and robustness to the gate protocols, in addition to the speed-up. To optimize the gate performance in this set-up, we need to apply optimization techniques  that deal not only with the temporal parameters of the laser but also with the spatial  position of the qubits or the field spatial structure. 

In the first  application of the spatio-temporal control of the qubits, we proposed the
symmetric orthogonal protocol (SOP), where all the {\em odd}-numbered pulses 
%in the sequence
as well as all the {\em even}-numbered pulses in the sequence, 
are time-delayed replicas \cite{Sola_Nanoscale2023}.
%\tcr{he quitatdo "of the same (odd or even) pulses"}
%have identical features.
The gate mechanism relied on the presence of a dark state in the Hamiltonian, for which 
%odd pulses and even pulses in the sequence, 
even and odd pulses had to be, in a certain sense, orthogonal to each other.
Then we developed a general mechanism analysis of the gates in terms of quantum pathways
and we showed that by optimizing the fields with fewer constraints, a plethora %canopy 
of different
optimal protocols with very high fidelities could be obtained, classified, and ranked,
%with some solutions showing  stricking correlations among the  field parameters.
according to their dynamics \cite{previouswork}. % the system.

In this work, we generalize our approach to treat different $2$-qubit and $3$-qubit
entangling gates in $N$-qubit systems.
%From the temporal point of view, we control the pulse areas and the sequence of non-overlapping pulses \cite{Rice_2000,Shapiro_2011,shore_2011,Malinovsky_PRA2004,Malinovsky_PRL2004,Malinovsky_PRL2006}.
%Control may be extended to other variables, including the time-delays and the relative phases between the pulses.\cite{Kosloff_PRA2003,Tesch_CPL2001,Tesch_PRL2002,Koch_PRA2008,Goerz_NJP2014, Caneva_PRA2011,Glaser_EPJD2015,Koch_EPJQT2022,Koch_PRA2011,Saffman_PRA2016}
%The spatial control can be achieved by different means. Two tentative implementations of the set-up are shown in Fig.\ref{scheme}. The most straight-forward one involves using a superposition of overlapping phase-locked Gaussian modes centered at different qubits, instead of a single field, for each pulse in the sequence.
%However, depending on the protocol found, it may be possible to use different modes of the electromagnetic field, or a linear combination of them\cite{Murty_AppOpt1964}, instead of the superposition of the Gaussian TEM$_{00}$ beams. In any case, it is necessary to dispose of sufficiently complex light structures\cite{Forbes_NPhoto2021,Rubinsztein_Dunlop_JOpt2016}.
%
Instead of searching and studying a single realization of the gate, 
{\it e.g.} the highest-fidelity protocol,
we use quantum
optimal control techniques to scan and characterize the full space of optimal
solutions \cite{Kosloff_PRA2003,Goerz_NJP2014,Caneva_PRA2011,Goerz_JPB2011}.
The rate of success of the algorithm over a very broad sample
of initial conditions is analyzed as a simple measure of the density (and quality) of solutions
for different constraints in the space of parameters.
We further study the mechanisms under which every optimal protocol 
over a broad family of pulse sequences operates
using quantum pathways, and rank the protocols following the procedure recently 
proposed in \cite{previouswork}.

Analyzing the overall patterns of the optimal protocols,
we seek to answer questions like: What is the minimal energy introduced 
through external fields necessary for the gates to operate? And how much per qubit?
Are all qubits used equally during the gate dynamics? How much does it
depend on the type of gate or on the number of qubits in the system?
And finally, can we infer which qubit arrangements or geometries are more 
promising for high fidelity gates? 

% PASAR A CONCLUSIONES:
%Using a simple semi-analytical approach, we develop procedures that allow us to select specific protocols within the dense manifold of solutions, as well as to indirectly influence the search in parameter space depending on the geometry of the qubits.
%Our results suggest that fast and reasonably robust solutions may be easier to find when not all distances between the qubits are equal, but on the other hand, the asymmetry can be limited to only one of the dimensions.
%These general features can be used as guides to propose simplified designs  targeting specific subsets of non-independent qubits.

% PASAR A CONCLUSIONES:
%Our study is a first approach on the subject of full quantum control by addressing both the spatial and temporal features of the lasers and assumes ideal conditions (a simplified Hamiltonian, zero temperature, no laser noise).
%In a first approximation with Rydberg atoms, working with fast gates justifies  neglecting most sources of decoherence. Preliminary calculations show a relatively weak dependence  of the fidelity on intensity fluctuations in the laser fields, as well as the atomic motion of the atoms, which can lead to a $\sim 1$\% loss of fidelity at $\sim25 \mu$K. More detailed studies will be needed to address the practical implementation of the protocols. 

\section{Platform setup and analytical Model}

\begin{figure}
\includegraphics[width=8.5cm]{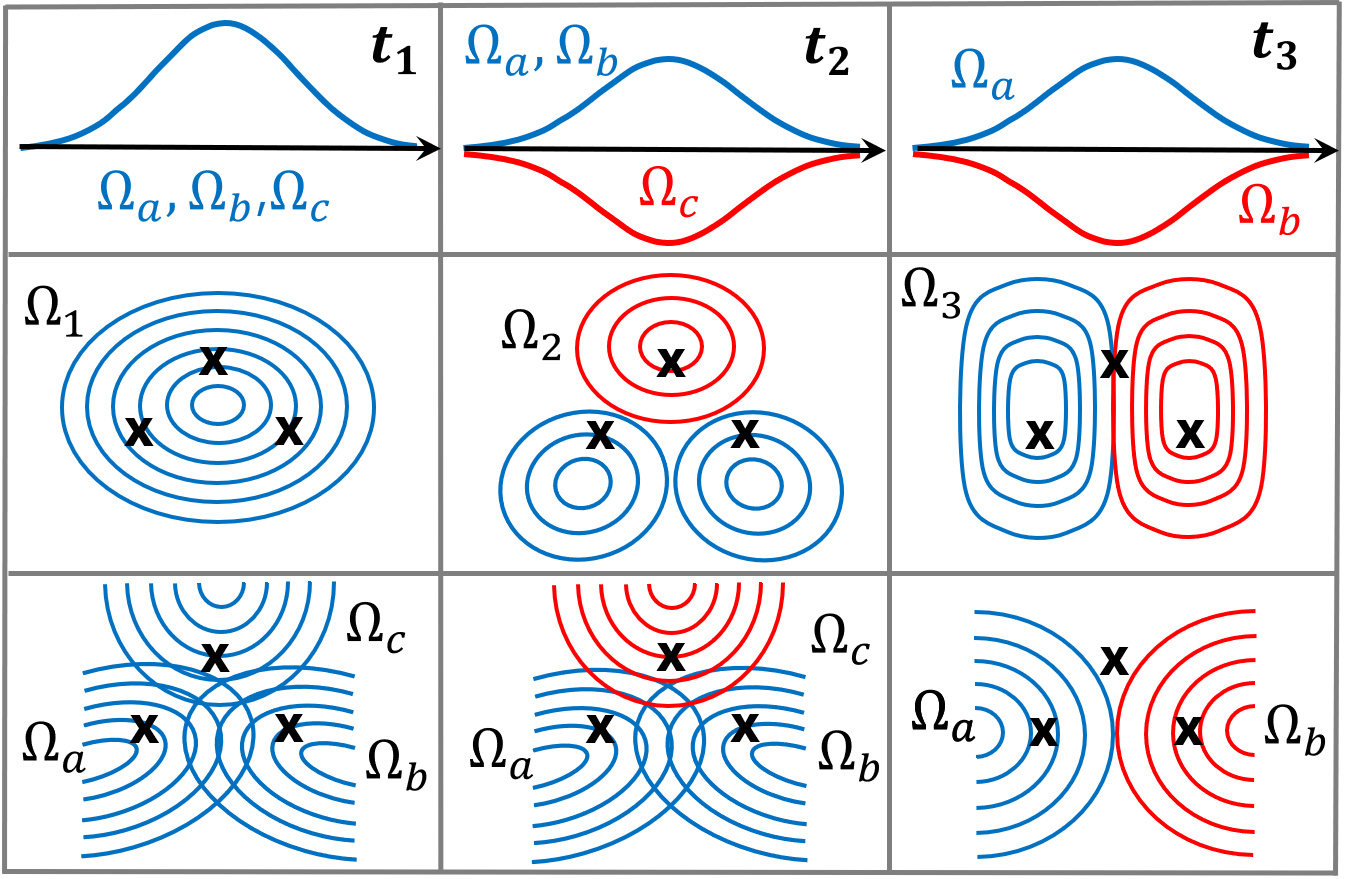} %{scheme-3q-c.png}
\caption{Scheme showing two possible implementations of the spatio-temporal 
control of trapped qubits, located at the $\times$ positions, 
using phase-locked Gaussian beams TEM$_{00}$ centered at each qubit, 
$\Omega_a, \Omega_b, \Omega_c$ (second row)
or structured pulses $\Omega_k$ whose spatial profile at the qubits
coincides with the linear combination of TEM$_{00}$ (third row).
The sequence of operations that governs the temporal evolution of the state 
depends on the pulse sequence, shown in the first row.}
\label{scheme}
\end{figure}

As a quantum platform for information processing, we
consider a set of closely separated atoms trapped by optical tweezers, 
where the computational states are encoded in low-energy hyper-fined states,
while the entangling gates (CZ or similar), which imply population return with
a sign flip conditional on the state of the control qubit, use the dipole-blockade
mechanism to gain a phase accumulation during the dynamics through a Rydberg state  of the atom.
With strong magnetic fields, the energy splitting between the qubit states can reach almost 
$\Delta \sim 10$ GHz \cite{Grangier_PRL2010,Scotto_phd2016}, while the energy difference between adjacent Rydberg states when the principal quantum number is around $n \sim 70$, 
is typically larger\cite{gallagher_1994}.  
Therefore, in setups where the atoms are close enough that the dipole-dipole interaction is on
the order of the energy splitting, $d_{\cal B} \sim  \Delta$, the gates can operate
in  the nanosecond time-scale.
The price to pay is the need to find protocols that are robust under the parallel excitation of several qubits.
In this work, we use the spatio-temporal control procedure to achieve precisely this goal. 
While the number of atoms in typical traps can easily reach the
hundreds, only a small subset can be excited by the same laser pulses at the same time.
They represent the minimal unit  that must be controlled to engineer the gate.
Here we consider subsets of $2$ to $4$ qubits.

In this work, we generalize the Hamiltonian and time-evolution operators to treat different entangling 
gates in $N$-qubit systems, under the same approximations used in \cite{Sola_Nanoscale2023} and \cite{previouswork}.
We model the effect of the field overlapping several qubits by
defining Rabi frequencies
$\tilde{\Omega}_{jk}(\vec{r}_j,t) = c_{jk} \mu_{0r} E_k(t)/\hbar = c_{jk}\Omega_k(t)$,
that depend on {\em geometrical factors}
$c_{j,k}$, which give the local effect of the field $\Omega_k(t)$ on the qubit $j$.
The geometrical factors can be partially incorporated into the Franck-Condon factor 
$\mu_{0r}$ so we can assume, without loss of generality, 
that $c_{jk}$  are normalized to one for each pulse,
($\sqrt{\sum_j^N c_{j,k}^2} = 1$ for all $k$). 
It is convenient to define the row vector 
${\bf e}_k^\intercal \equiv \langle {\bf e}_k| = (e_{1k}, e_{2k},\ldots, e_{nk})$
(${\bf e}_k \equiv |{\bf e}_k\rangle$ is the column vector in bracket notation) 
formed by all the $e_{i,k}$ geometrical factors of a given pulse. 
We will call ${\bf e}_k$ the {\em structural vectors}.
From the temporal point of view,
we control $\Omega_k(t)$. In this work only the pulse areas and the length of the sequence of non-overlapping
pulses will be important \cite{Rice_2000,Shapiro_2011,shore_2011,Malinovsky_PRA2004,Malinovsky_PRL2004,Malinovsky_PRL2006}.
%In this work $\Omega_k(t)$ are real, so the relative phase between the pulses is fixed as either $0$ or $\pi$.
Control may be extended to other variables, including the time-delays and the relative phases
between the pulses \cite{Kosloff_PRA2003,Tesch_CPL2001,Tesch_PRL2002,Koch_PRA2008,Goerz_NJP2014,
Caneva_PRA2011,Glaser_EPJD2015,Koch_EPJQT2022,Koch_PRA2011,Saffman_PRA2016}.

The spatial control is encoded in ${\bf e}_k$ and can be achieved by different means.
In \cite{Sola_Nanoscale2023} we proposed the use of hybrid modes of light to
allow a wide range of possible values for ${\bf e}_k$, including negative amplitudes. 
A possible generalization for spatially non-orthogonal pulses may
require more complex structured light\cite{Murty_AppOpt1964}, 
such as those sketched in Fig.\ref{scheme} (second row).
A simpler laboratory implementation, shown in the third row, can be achieved using
a superposition of overlapping phase-locked Gaussian modes centered at different qubits, 
instead of a single field, for each pulse in the sequence.
In any case, it is necessary to dispose of sufficiently complex light
structures to extend the control to the spatial domain \cite{Forbes_NPhoto2021,Rubinsztein_Dunlop_JOpt2016}.
A different alternative could be achieved by moving the atoms to different positions, but this would
slow down significantly the gate duration.
%Two tentative implementations of the set-up are shown in Fig.\ref{scheme}. The most straight-forward one involves using a superposition of overlapping phase-locked Gaussian modes centered at different qubits, instead of a single field, for each pulse in the sequence.
%However, depending on the protocol found, it may be possible to use different modes of the electromagnetic field, or a linear combination of them\cite{Murty_AppOpt1964}, instead of the superposition of the Gaussian TEM$_{00}$ beams.In any case, it is necessary to dispose of sufficiently complex light structures\cite{Forbes_NPhoto2021,Rubinsztein_Dunlop_JOpt2016}.

For each qubit, there are $2$ states that form the computational basis, which are the states
that can be initially populated, and an additional Rydberg state, so the full Hilbert space can be spanned by $3^N$ states.\cite{nielsen_chuang_2010}
In the strong dipole-blockade regime, where only
a single qubit can be excited to a Rydberg state,
the number of states that can be populated during the dynamics is $2^N + N\, 2^{N-1}$,
forming $2^N$ disconnected systems (the Hamiltonian is a block matrix under
the approximations considered in the model), that
can be classified as:

\begin{itemize}

\item A $V^{(N,1)}$ subsystem formed by the ground state $|0\cdots 0\rangle$ coupled to the $N$ Rydberg 
excitations $|0\cdots r_e \cdots 0\rangle$ by $c_{e,k}\Omega_{k}$, where $e$ can occupy the
positions $1$ to $N$ ($1\le e \le N$). 

\item $N$ $V^{(N-1,m)}$ subsystems 
formed by the single-excited qubit states $|0\cdots 1_h \cdots 0 \rangle$ ($1\le h \le N$), each
coupled to the $N-1$ Rydberg excited states $|0\cdots r_e \cdots 1_h \cdots  0 \rangle$
 ($e \ne h$) by $c_{e,k}\Omega_k$. 
The second index $m$ distinguishes the $N$ different $V^{(N-1,m)}$ structures
which differ by the states that are participating. 
We order them choosing the index $m$ as the excited qubit $h$, so that 
for $m=1$ the ``ground'' state is $|10\ldots 0\rangle$
and the remaining $N-1$ states are $|1r\ldots 0\rangle, \ldots |10\ldots r\rangle$.

\item 
$N! / ( N_e!(N-N_e)! ) $ $V^{(N-N_e,m)}$ subsystems formed by the $N_e$-excited qubit states 
coupled to their possible Rydberg excitations.

\item $\vdots$

\item $N$ two-level systems ($V^{(1,m)}, m \in [1,N]$) with $|1\cdots 0_e \cdots 1\rangle$ and $|1\cdots r_e \cdots 1\rangle$

\item The uncoupled $N$-excited qubit state $|1\cdots 1\rangle$ ($V^{(0,1)}$ system).
\end{itemize}

We can treat the dynamics of each subsystem independently,
but the outcome must be conditioned to the
logic of the entangling $2$ (or $3$) qubit gate.
We consider the following version of  the CZ gate, ${\cal P}_{ab}$, acting on qubits
$a$ and $b$  ($a$ is by definition the first qubit, $b$ the second)
with logic  tableaux
$|00\cdots\rangle \rightarrow -|00\cdots\rangle$,
$|01\cdots\rangle \rightarrow -|01\cdots\rangle$,
$|10\cdots\rangle \rightarrow -|10\cdots\rangle$,
$|11\cdots\rangle \rightarrow |11\cdots\rangle$,
regardless of the values of any additional qubits $c$, $d$, etc.
%(This C-PHASE type gate changes the signs of the operations in the usual definition of a C-PHASE or CZ gate, but this change can be absorbed into a global phase.)
The set of conditions can be summarized in the diagonal elements of the matrix ${\sf P}_{ab}$.
For instance, in a $3$-qubit system, the ${\cal P}_{ab}$ has the
signature $\mathrm{diag}\lbrace -1, -1, -1, -1, -1, -1, 1, 1\rbrace$ for a basis ordered
as $|000\rangle, |001\rangle, |010\rangle, |100\rangle, |011\rangle, |101\rangle, |110\rangle, |111\rangle$.
In addition to studying the optimization of the ${\cal P}_{ab}$
gate, we also consider $3$-qubit entangling gates
as ${\cal P}_{abc}$,
where the matrix ${\sf P}_{abc}$ is diagonal with signature
$\mathrm{diag}\lbrace -1, -1, -1, -1, -1, -1, -1, 1 \rbrace$.
%with the basis ordered as $\lbrace |000\rangle, |010\rangle, |100\rangle, |001\rangle, |101\rangle, |011\rangle, |110\rangle, |111\rangle \rbrace$.

For non-overlapping pulses with single-photon Rabi frequency $\Omega_k(t)$, 
in the rotating-wave approximation and in the interaction
picture, the Hamiltonian is a direct sum of subsystem $V^{(n,m)}$ Hamiltonians, 
${\sf H}_k^{(n,m)}(t)$, 
which are everywhere zero except for the first row/column, with matrix elements
$H_{k,11}^{(n,m)} = 0$ and
$H_{k,1j}^{(n,m)}(t) =  -c_{j-1,k}\Omega_k(t)/2$ ($j=2,N$),
where $t$ is defined within the domain of the pulse $k$.

For each $V^{(n,m)}$, the time-evolution operator is a $(n+1) \times (n+1)$ 
matrix that  can be written as a time-ordered
product of the time-evolution operators for each pulse in the sequence,  
$${\sf U}_T^{(n,m)} = \prod_{k=0}^{N_p-1} U^{(n,m)}_{N_p-k}$$
where %${\sf U}^{(n,m)}_{k} = $
\begin{widetext}
\begin{equation}
{\sf U}^{(n,m)}_{k} =
\left( \begin{array}{ccccc}
\cos S^{(n,m)}_k & i e_{1,k} \sin S^{(n,m)}_k & i e_{2,k} \sin S^{(n,m)}_k 
& \cdots & i e_{n,k}\sin S^{(n,m)}_k \\ \\
i e_{1,k} \sin S^{(n,m)}_k & 1 + e_{1,k}^2 \left[ \cos S^{(n,m)}_k - 1 \right] & 
e_{1,k}e_{2,k} \left[ \cos S^{(n,m)}_k - 1 \right] & \cdots & 
e_{1,k}e_{n,k} \left[ \cos S^{(n,m)}_k - 1 \right] \\
\vdots & \vdots & \vdots & \ddots & \vdots \\   % iddots
i e_{n,k} \sin S^{(n,m)}_k & e_{n,k}e_{1,k} \left[ \cos S^{(n,m)}_k - 1 \right]  & 
e_{n,k}e_{2,k}  \left[ \cos S^{(n,m)}_k - 1 \right] & \cdots & 1 + e_{n,k}^2 
\left[ \cos S^{(n,m)}_k - 1 \right] \end{array} \right)
\end{equation}
\end{widetext}
where the geometrical factors,
$e_{j,k} = c_{j,k} / f_k^{(n,m)}$, are normalized with 
 \begin{equation}
f_k^{(n,m)} = \sqrt{\sum_{i \in m}^n c_{i,k}^2} \ ,
\end{equation}
that depend on the subsystem through the coefficients that enter in $f_k^{(n,m)}$.
The general form of the propagator is, however, independent of the $m$ index. 
%This index only specifies the elements of the basis to which the propagator is applied.
If $n =1$, $c_{1,k}$ can only be $\pm 1$, whereas if $n=N$, $f_k^{(N,1)}=1$, due to
the normalization of the $c_{jk}$ geometrical factors.
The mixing angles are
\begin{equation}
S_k^{(n,m)} = \frac{1}{2} f_k^{(n,m)} \int_{-\infty}^{\infty} \Omega_k(t) dt
= \frac{1}{2} f_k^{(n,m)} A_k \ ,
\end{equation}
where $A_k$ are the pulse areas.

%\section{Optimal protocols for different entangling gates}
\section{Results and Analysis}

Using the Nelder and Mead simplex optimization scheme 
with linear constraints \cite{Nelder_CJ1965,Powell_AcNum1998}, 
we optimize the pulse areas $A_k$ ($k \in [1, N_p]$),
and the geometrical parameters $e_{jk}$ ($j \in [1, N], k \in [1, N_p]$), %geometrical parameters, 
to maximize
the fidelity of the gate.
In this work $\Omega_k(t)$ are real, so the relative phase between the pulses is fixed as either $0$ or $\pi$.
The algorithm is applied to ${\cal N}_T = 10^5$ different initial configurations of
the parameters obtained through a uniform distribution within some chosen range.
The geometrical factors are constrained such that a minimum value of $|e_{jk}| 
\ge \sigma$ is imposed. Protocols with smaller $\sigma$ accept
solutions where the influence of the pulse on both qubits at the same time can
be smaller, which are related to more separated qubits (or pulse beams with a wider beam waist).
We also perform optimizations forcing the positivity of
the geometrical factors, $e_{jk} \ge \sigma$ (p-restricted protocols), 
which we denote by $\sigma^+$. 

\begin{figure}
\hspace*{-0.5cm}
\includegraphics[width=7cm]{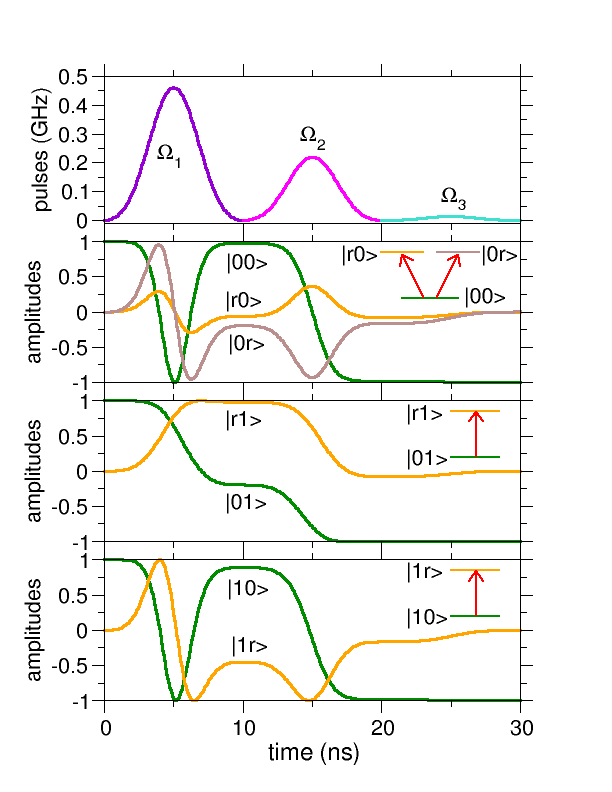}
\caption{A particular $3$-pulse optimal protocol for the ${\cal P}_{ab}$ gate
with $\sigma = 0.3$. We show the pulses and amplitudes for the dynamics starting
in the different computational basis. The amplitudes in the starting states are 
real while in the ancillary states are always purely imaginary.}
\label{Dyn}
\end{figure}

Fig.(\ref{Dyn}) shows the pulses and population dynamics starting from the different
computational basis for one optimal protocol in a $2$-qubit system using three pulses, 
with $\sigma = 0.1$, which gives an infidelity
($\epsilon = 1-F$) of $7\cdot 10^{-7}$ 
for the ${\cal P}_{ab}$ gate.
In this particular protocol, with an overall pulse area $A_T = A_1 + A_2 + A_3 = 5.8\pi$, 
%A1 = 3.87pi, A2 = 1.84pi, A3 = 0.12pi
% a1 = 0.3, a2 = 0.5, a3 = -0.43 (b2 and  b3 < 0), e1*e2 = -0.67, e2*e3 = 0.57, e1*e3 = -0.98
the third pulse is basically used to correct the fidelity of a $2$-pulse sequence.
The first pulse acts mainly on the second qubit ($e_{a1}^2 = 0.09$), while the second and third
act on the first qubit ($e_{b2}^2 = 0.25$, $e_{b3}^2 = 0.17$).
%\tcr{first qubit es 'a' y second qubit es 'b'??}
The square of the 
geometrical factor in the least used qubit measures the degree to which the pulse acts on
more than one qubit and hence is an indication of how much the protocol relies on
interdependent qubits. The scalar products $\langle e_1 | e_2 \rangle = -0.67$,
$\langle e_2 | e_3 \rangle = 0.57$, $\langle e_1 | e_3 \rangle = -0.97$ indicate
the correlation among the structural vectors. For this strategy, $\Omega_1(\vec{r},t)$
and $\Omega_3(\vec{r},t)$, almost revert their role from the spatial point of view.
The first pulse induces a $4\pi$ transition from $|00\rangle$,  a $\pi$ transition from $|01\rangle$
and a $2\pi$ transition from $|10\rangle$. Only the $|01\rangle$ state goes to the ancillary state
$|r1\rangle$ at the end of the first pulse. The second pulse is responsible for a 
$2\pi$ transition from $|00\rangle$ and $|01\rangle$ and a $\pi$ transition from $|r1\rangle$.
%According to the mechanism classification proposed in \cite{previouswork}, the protocol is a 0-loop for the $V^{(2,1)}$ and $V^{(1,2)}$ subsystems, and a 1-loop for the $V^{(1,1)}$ subsystem.

\begin{figure}
\hspace*{-0.5cm}
\includegraphics[width=7.5cm]{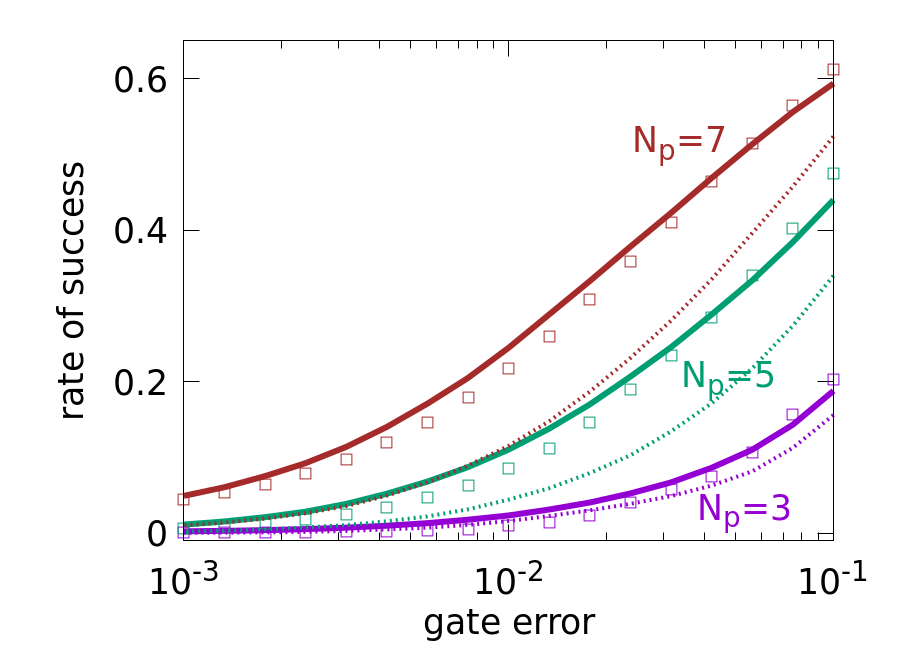}
\caption{Rate of success of the optimization as a function of the gate, ${\cal P}_{ab}$ (solid lines) and  ${\cal P}_{abc}$ (squares), for different pulse sequences, imposing $\sigma = 0.1$. In dots we show the result for the ${\cal P}_{ab}$ gate imposing  $\sigma^+ = 0.1$.}
%\includegraphics[width=9cm]{performance-Pab_Pabc-mod2.png}
%\caption{(a) Rate of success of the optimization as a function of the gate, ${\cal P}_{ab}$ (solid lines) and  ${\cal P}_{abc}$ (squares), for different pulse sequences, imposing $\sigma = 0.1$. In dots we show the result for the ${\cal P}_{ab}$ gate imposing  $\sigma^+ = 0.1$.
%(b) Rate of success of the optimization for the ${\cal P}_{ab}$ gate using $6$-pulse sequences in systems with different numbers of qubits. In solid lines, we show the result when $\sigma = 0.1$. We use squares for results with $\sigma = 0.3$ but only forced on qubits $a$ and $b$. The field can take any value on the remaining qubits. Dashed lines show the results when $\sigma = 0.3$ in any  (not initially assigned) two qubits.}
\label{perP}
\end{figure}

But rather than studying individual results, we want  to focus on general trends,  for which
we analyze common features of the set of all optimal protocols.
One of the most interesting conclusions can be obtained by studying the rate of success
of the algorithm, which refers to the percentage of initial conditions
(${\cal N}_\epsilon/{\cal N}_T$)
%that is, the percent of initial conditions ${\cal N}_\epsilon/{\cal N}_T$ 
that lead to optimal gates with infidelity smaller than a certain threshold $\epsilon$.
The specific curves may change slightly depending on the set of
initial conditions,
%for different sets of initial conditions,
so it is better to compare the curves using a single, very large set.

Fig.\ref{perP} shows the rate of success for gates ${\cal P}_{ab}$ and ${\cal P}_{abc}$
for different pulse sequences in $3$-qubit systems, where we impose $\sigma = 0.1$.
Larger pulse sequences use more variational parameters and as expected, have higher rates of success,
but one can always find protocols with fidelity greater %larger 
than $0.99$ using only %with 
$2$-pulse sequences
(or $3$-pulse sequences in $4$-qubit systems).
%  BUT THERE ARE ALWAYS SOLUTIONS
When the number of parameters becomes too large ($N_p \gtrsim 8+N$) the algorithms do not give
significantly better results.
The overall behavior is similar for both gates, but in general, there are more successful 
protocols with low fidelity for the ${\cal P}_{abc}$, particularly when the number of pulses
increases, while  the  opposite is true at the high-fidelity limit.
Enforcing positive geometrical factors leads to a decay in the rate of success
especially in longer pulse sequences, but this decay is much steeper in the ${\cal P}_{abc}$ gate
in $3$-pulse sequences.
This poses interesting questions concerning the main mechanisms used %explored 
by the different protocols for the different gates.

\begin{figure}
\includegraphics[width=8.6cm]{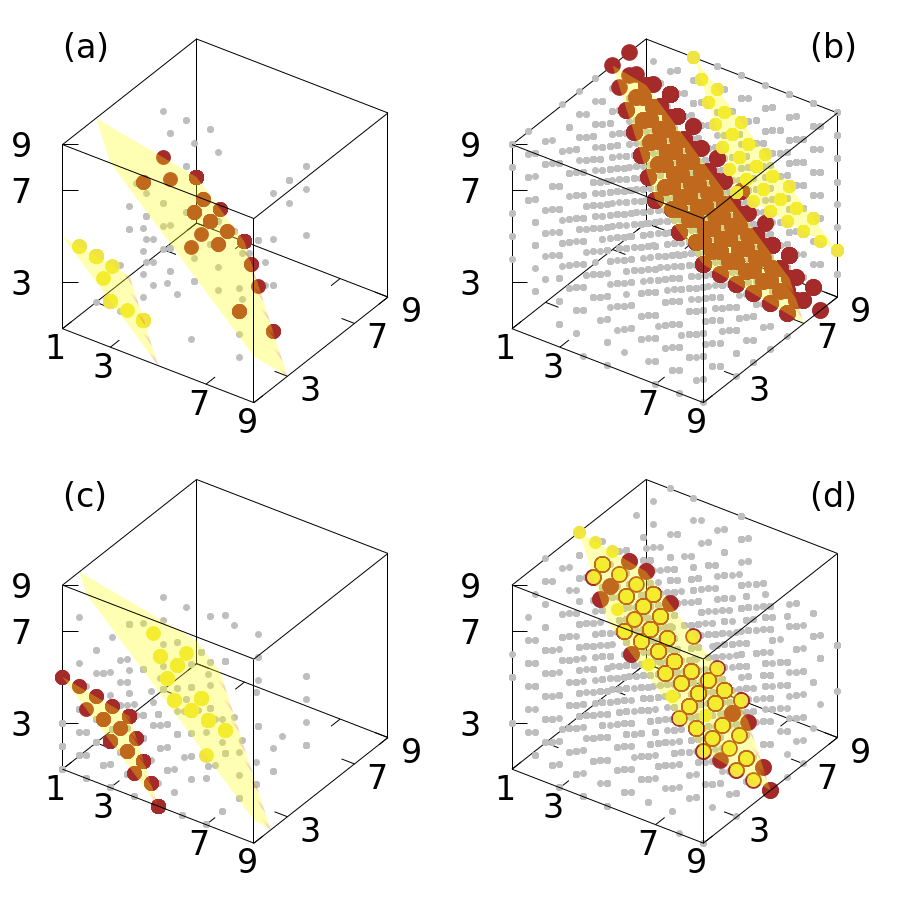}
\caption{Diagram
%m-cube 
showing the most frequent mechanisms for the two-level subsystems 
$V^{(1,1)}, V^{(1,2)}, V^{(1,3)}$ (top row) and the $V$ subsystems
$V^{(2,1)}, V^{(2,2)}, V^{(2,3)}$ for the ${\cal P}_{ab}$
gate (brown circles) and the ${\cal P}_{abc}$ gate (yellow circles).
All used mechanisms  for the ${\cal P}_{ab}$ gate are shown with gray circles.
The first column shows  results for $3$-pulse sequences; the second, for
$5$-pulse sequences. The circles have been made of slightly
different sizes so that common mechanisms for both gates are mixed brown-yellow circles.}
\label{cube3q}
\end{figure}

%\begin{figure}
%\includegraphics[width=7.0cm]{mechanism1.png}
%\caption{.}
%\label{mech}
%\end{figure}
\begin{figure}
\includegraphics[width=8.6cm]{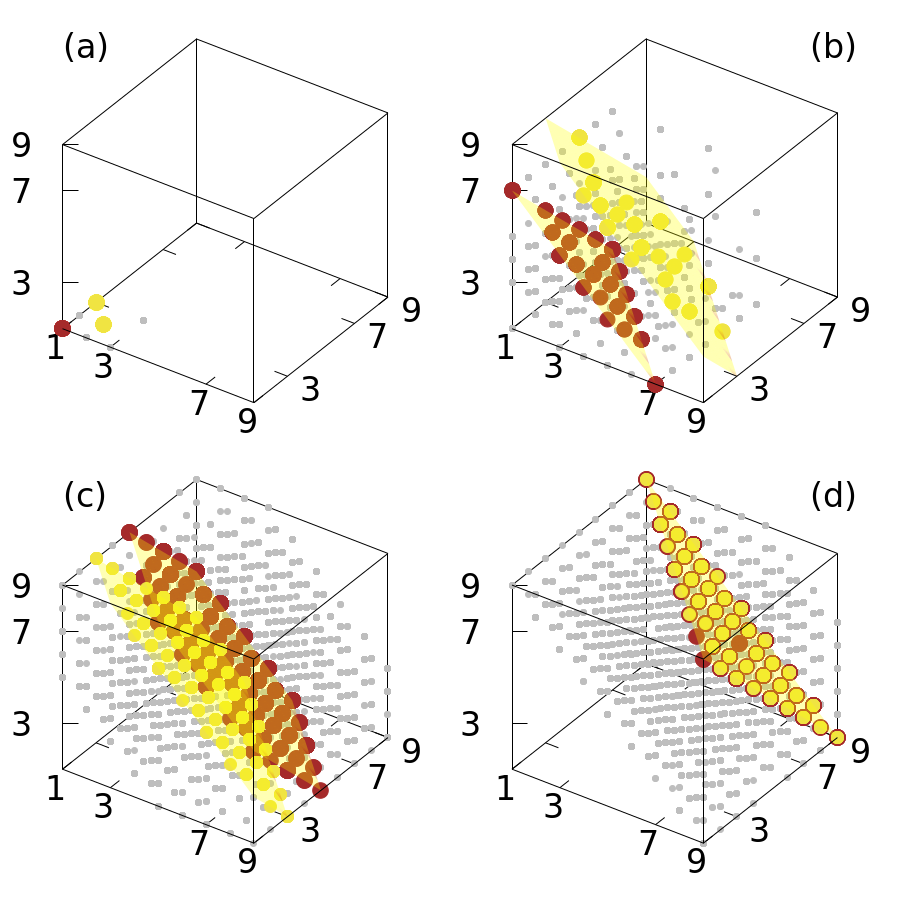}
%m-cube showing the most frequent mechanisms for the ${\cal P}_{ab}$ gate (brown circles) and the ${\cal P}_{abc}$ gate (yellow circles) and all observed mechanisms (gray) used by the optimal protocols using (a) 2-pulses, (b) 3-pulses, (c) 4-pulses and (d) 5-pulse sequences. The circles have been made of slightly different \textcolor{blue}{sizes} so that common mechanisms for both gates are mixed brown-yellow circles.
\caption{Diagram
showing the most frequent mechanisms for the for the ${\cal P}_{ab}$ gate (brown circles) and the ${\cal P}_{abc}$ gate (yellow circles) and all observed mechanisms (gray) used by the optimal protocols using (a) 2-pulses, (b) 3-pulses, (c) 4-pulses and (d) 5-pulse sequences. The circles have been made of slightly different sizes so that common mechanisms for both gates are mixed brown-yellow circles.}
\label{cube3qfull}
\end{figure}

To classify and visualize the mechanisms of  the protocols, we  use our recently
proposed procedure based on quantum pathways \cite{previouswork}. %on $2$-qubit systems
For each starting subsystem $V^{(n,m)}$, we can write the matrix element of the time evolution operator,
$U_{T,11}$ as a contribution of $0$-loops (the state of the system after each pulse is the initial
state), $1$-loops (the population flops to the Rydberg state
after one pulse and returns after the next one), $d$-loops (the population stays in
the Rydberg state during the action of one or more consecutive pulses, acting like the $0$-loop 
for the Rydberg states) and $2$-loops (the population cycles two times through the excited state).
As a reference, the protocol dynamics shown in Fig.\ref{Dyn} is a 0-loop for the 
$V^{(2,1)}$ and $V^{(1,2)}$ subsystems, and a 1-loop for the $V^{(1,1)}$ subsystem.
For pulse sequences with $N_p \le 5$, we do not need to consider extra loops, so 
$U^{(n,m)}_{T,11} = u^{(n,m)}_0 + u^{(n,m)}_1 + u^{(n,m)}_d + u^{(n,m)}_2$,
where the expressions for the particular matrix  elements of  the time  evolution operators,
obtained in \cite{previouswork}, are valid for $N$-qubit systems.
Finally, to represent the results in the most simple way, we first define the
coordinates of a point in a square for each subsystem $V^{(n,m)}$, 
\begin{eqnarray}
x^{(n,m)} = u_0^{(n,m)} + u_1^{(n,m)} - u_d^{(n,m)} - u_2^{(n,m)} \nonumber \\
y^{(n,m)} = u_0^{(n,m)} + u_d^{(n,m)} - u_1^{(n,m)} - u_2^{(n,m)} 
\end{eqnarray}
and then partition each square into $9$ boxes, ranking the mechanism as
a number $\omega^{(n,m)} \in [1,9]$ depending on the box where $(x^{(n,m)},y^{(n,m)})$ 
is located
\footnote{Defining the floor integers (greatest integer smaller than the real number)
$\lfloor {x}^{(n,m)}\rfloor = \lfloor l (x^{(n,m)}+1)/2 \rfloor +1$,
$\lfloor {y}^{(n,m)} \rfloor = \lfloor l (y^{(n,m)}+1)/2]+1 \rfloor$ 
(where $l =3$ is the number of divisions of each m-square side,
$\lfloor {x}^{(n,m)}\rfloor, \lfloor {y}^{(n,m)}\rfloor \in [1,3]$), we call
$\omega^{(n,m)} = \lfloor {y}^{(n,m)}\rfloor + l( \lfloor {x}^{(n,m)} \rfloor-1 )$
the number that ranks the mechanism for each $V^{(n,m)}$.}.
Pure or dominant 0-loops correspond to $\omega = 1$, 1-loops to 
$\omega=3$, d-loops to $\omega=7$, and 2-loops to $\omega=9$.
In between, $\omega$ ranks collaborative mechanisms among the closest pure mechanisms, 
or possibly fully collaborative mechanisms, as is the case of $\omega = 5$.
%  for Np = 2 w in 1,3; Np = 3, w in 1,7; Np =4,5 w in 1,9

For a $3$-qubit system, for which there are $7$ subsystems 
(plus the decoupled $|111\rangle$ state),
we obtain $7$ coordinates.
The signature of the different entangling gates differs only on the
two-level systems, so we expect the prevalent mechanisms for the
${\cal P}_{ab}$ and ${\cal P}_{abc}$ gates to differ %be different 
mainly in the dynamics of the two-level subsystems. 
In Fig.\ref{cube3q} we represent the mechanism of each protocol with %by 
a point in a cube corresponding to %formed by the set of mechanisms of the two-level subsystems, 
$\left( \omega^{(1,1)}, \omega^{(1,2)}, \omega^{(1,3)} \right)$ for
pulse sequences with $3$ (a) and $5$ pulses (b), where we have selected
protocols with fidelity higher than $0.99$ with $\sigma = 0.1$.
Although protocols with larger pulse sequences explore all possible mechanisms
(shown as gray points in Fig.\ref{cube3q} for the ${\cal P}_{ab}$ gate)
the most prevalent mechanisms are clearly different for the two  gates
(shown as brown points for ${\cal P}_{ab}$ and yellow points for ${\cal P}_{abc}$)
as they never overlap. These sets vary little for different constraints
($\sigma = 0.3$, p-constrained mechanisms) or different error thresholds
($\epsilon > 10^{-3}$), but depend strongly on the number of pulses.

For $3$-pulse sequences, the most used mechanisms lie on
a single plane for ${\cal P}_{abc}$,
with $\omega^{(1,1)}+ \omega^{(1,2)} + \omega^{(1,3)} = 7$.
In the  ${\cal P}_{abc}$ gate, all two-level subsystems play the same role, 
and this symmetry is passed on to the $\omega^{(1,n)}$ which can be
interchanged.
The total value $\omega_T^{(1)} = 7$ shows preference for 0-loops and 1-loops
or their superposition (but not all the mechanisms within the plane are used equally).
This is not the case for the  ${\cal P}_{ab}$ gate.
In the latter, not all dominant mechanism lie on a single plane.
In addition, $\omega_T^{(1)}$ is much larger for the ${\cal P}_{ab}$ gate,
showing preference of d-loops.
However, for $5$-pulse sequences the prevalent mechanisms for the 
${\cal P}_{abc}$ gate lie on two planes, while those of the ${\cal P}_{ab}$ lie
on a single plane. Still, larger $\omega_T^{(1)}$ are used in the ${\cal P}_{ab}$
than in the ${\cal P}_{abc}$ gate.

The different mechanisms needed to achieve high fidelity protocols
used in the $V^{(1,n)}$ subsystems
influence the dominant mechanisms explored by the other subsystems as well.
In Fig.\ref{cube3q}(c) and (d) we show the m-cube
formed by the set of mechanisms of the $V$ subsystems,
$\left( \omega^{(2,1)}, \omega^{(2,2)}, \omega^{(2,3)} \right)$ for
$3$ and $5$ pulse protocols. While the difference between ${\cal P}_{ab}$
and  ${\cal P}_{abc}$ protocols is obvious for $3$-pulse sequences to
the point of no-overlap in the distributions, the differences tend to disappear for
longer pulses sequences. %Interestingly, here $\omega_T^{(1)}$ is smaller for P_ab!
Again, most mechanisms lie on one or two planes.
The distribution is also different for the preferred mechanisms in
$V^{(3,1)}$ for $3$-pulse sequences, where $\omega^{(3,1)} = 1$ in ${\cal P}_{ab}$, 
while $\omega^{(3,1)} = 7$ in ${\cal P}_{abc}$.
However, the most used mechanisms tend to coincide again
for $5$-pulse sequences ($\omega^{(3,1)} = 7$).
In general, as the number of pulses increases, the set of mechanisms increases
and differences between the gates are less pronounced.

The mechanism analysis for the whole system is difficult to visualize in a single
plot for $3$-qubit systems, as we need $7$ coordinates to characterize every mechanism. 
%now $7$ subsystems for which we can obtain the m-square. 
We again observe correlations between the set of values for the different kinds
of subsystems.
%of $\omega^{(2,j)}$ and $\omega^{(1,j)}$ ($j \in [1,3]$).
To simplify the analysis, in Fig.\ref{cube3qfull} we form a cube with 
$\omega^{(3,1)}$ in the $xy$ plane, the projection of all $\omega^{(2,j)}$ in the $yz$ plane, and the projection of all $\omega^{(1,j)}$ in the $xz$ plane. 
The dominant mechanisms in ${\cal P}_{ab}$ (brown circles) and ${\cal P}_{abc}$ (yellow circles) are shown along the set of other less used mechanisms (gray circles, for
a frequency of $30$\% the most dominant one) 
for different pulse sequences and $\sigma = 0.1$.
%Fig.\ref{cube3q} %and Fig.\ref{cubecq} mechanisms shows the dominant mechanisms for ${\cal P}_{ab}$ (brown circles) and ${\cal P}_{abc}$ (yellow circles) along the set of other used mechanisms (gray circles) for different pulse sequences and $\sigma = 0.1$.
%

Again, most dominant solutions in both gates line in some particular planes,
showcasing the surprising symmetry where the preferred optimal protocols use 
the same mechanisms regardless of the initial state of the computational basis.
That is, if one can find a high fidelity protocol where the dynamics follows
a $0$-loop from $|000\rangle$ and $|001\rangle$ (or any other subsystem with
one excited qubit) and a $d$-loop starting from $|011\rangle$ (or any other
subsystem with two excited qubits), then it is likely that another high fidelity
protocol can be found when the dynamics follows a $d$-loop starting from $|000\rangle$
or $|010\rangle$, while all the other subsystems follow a $0$-loop.

Characteristically, we observe that 
$\omega_T = \omega^{(3,1)} + \omega^{(2,j)} + \omega^{(1,j)}$, with $j = 1, 2, 3$
increases with the number of pulses in the sequence, which correspond to
favoring 2-loops over d-loops, and d-loops over 1-loops in the collaborative mechanisms, 
as one moves from $3$ to $5$-pulse sequences.
The same behavior was observed studying the mechanism of two-qubit systems 
\cite{previouswork}.
We also observe that the set of preferred mechanisms is very different for
the ${\cal P}_{ab}$ and ${\cal P}_{abc}$ gates for short sequences, so they
do not share any dominant mechanism for $N_p = 2, 3, 4$. However, the planes
approach as the number of pulses increases, and fully overlap for $N_p = 5$.
%Since $\omega_T$ is odd, 2 pure + 1 collaborative, or 3 collaborative but "even" (2, 4, 6, 8)
%the ${\cal P}_{ab}$ are included in the same planes as in the 2-qubit systems, where now $\omega^{(3,1)} + \omega^{(2,j)} + \omega^{(1,j)} = \omega_T$, with $j = 1, 2, 3$. However, many mechanisms lying in the plane are not used (or not frequently used) in the 3-qubit systems.
%The behavior is very different in the ${\cal P}_{abc}$ gate for short sequences from the mechanistic point of view, so much so that the dominant  mechanisms practically do not overlap.  But as $N_p$ increases, the most used mechanisms approach and practically overlap for $5$-pulse sequences.

\begin{figure}
\includegraphics[width=7.5cm]{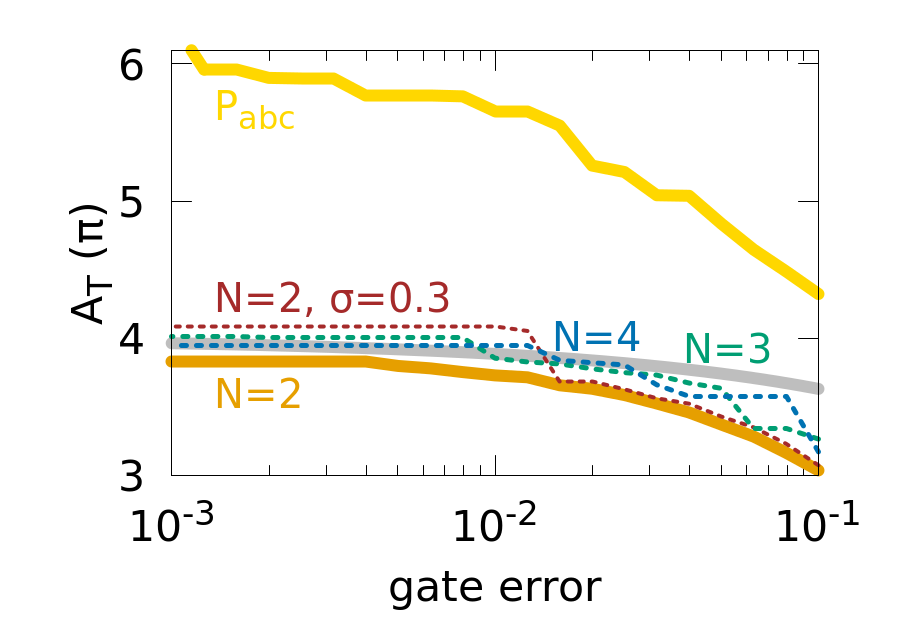}
\caption{Minimal pulse area used in the optimal protocols for different gate errors for
the ${\cal P}_{ab}$ gate implemented in $2$, $3$, and $4$ qubit systems, and
for the ${\cal P}_{abc}$ gate implemented in $3$ qubit systems.
For comparison, we also show in the gray line the theoretical minimal area required
for independent qubits.}
\label{minarea}
\end{figure}

To evaluate the minimal energy input necessary to operate the gates, we have
measured the total pulse area, $A_T = \sum_k A_k$, used in each protocol.
We performed optimizations constraining $A_T$ to be smaller than
a certain threshold, which was made as small as possible.
%, but we did not penalize the areas to give more flexibility. 
Fig.\ref{minarea} shows the minimal pulse areas found for the optimal protocols
with infidelity smaller than a certain gate error over a large set of initial conditions 
($\sim 10^6$), including different pulse sequences and values of $\sigma$.
%including different pulse sequences, values of $\sigma$, etc.

For independent qubits, Jaksch protocol for the CZ gate (${\cal P}_{ab}$) 
requires a minimal area of $A_T = 4\pi$ for perfect fidelity \cite{Jaksch_PRL2000}. 
The gray line in Fig.\ref{minarea} gives the theoretical value for different fidelities. 
Using non-independent qubits, the minimal energy can be reduced to $3.8\pi$ for
high fidelities (better than $0.999$) or $3\pi$ for low fidelities  (better than $0.9$).
This implies an energy reduction from $3.6$\% to more than $16$\% with respect
to independent qubits.
Strengthening the constraints (e.g. for closer qubits, $\sigma = 0.3$,
or demanding positive geometrical factors) implies an increase in the minimal area 
for high fidelity protocols of no more than $7$\%.
The extra energy is also needed in $3$ or $4$ qubit systems, but this cost goes  to
zero for lower fidelity gates.

On the other hand, the dominant mechanisms for the ${\cal P}_{abc}$ gate are completely
different as the gate requires a minimal energy of roughly $6\pi$ for high fidelity
protocols (around $2\pi$ per qubit, as the sign flip demands Rabi cycling).
However, for non-independent qubits at low fidelities, the minimal energy can be
substantially reduced. We have obtained protocols with at least $0.9$ fidelity with
$A_T = 4.3\pi$. While the minimal total areas do not depend on the number of pulses
in the ${\cal P}_{ab}$ gate, in the case of ${\cal P}_{abc}$,
shorter sequences ($N_p \lesssim 4$) perform with an energy penalty.

%\section{Control with asymmetric constraints}

\begin{figure}
\includegraphics[width=7.5cm]{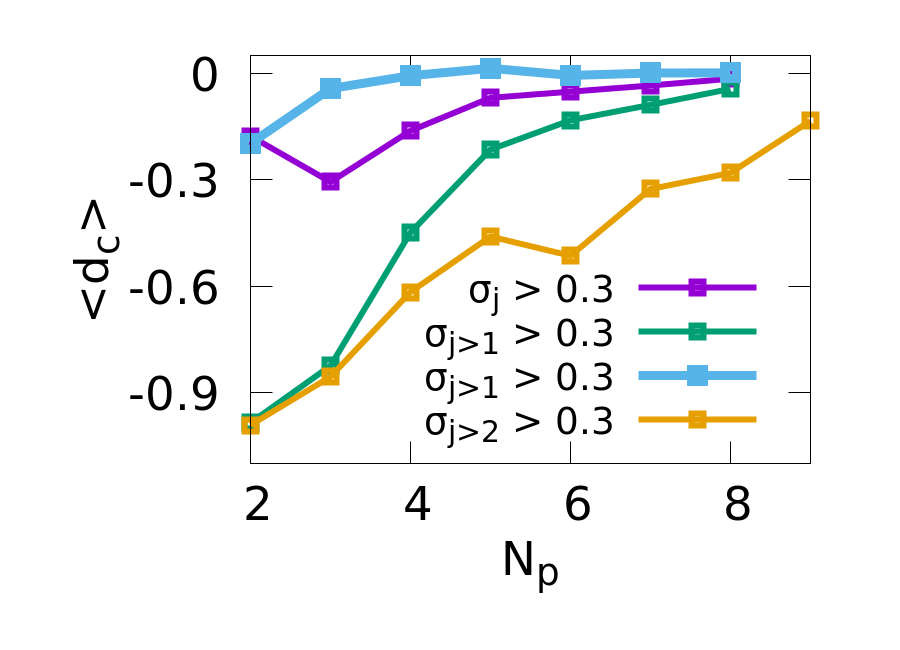}
\caption{Average contribution of qubit $c$ %or qubit $c$ plus $d$
in optimal protocols for systems with more than $2$-qubits 
as a function of the number of pulses in the sequence. 
We show the relative deviation from the uniform contribution, 
$N e_\alpha^2 - 1$. 
The blue line is for the ${\cal P}_{abc}$ gate, 
all the others for the ${\cal P}_{ab}$. 
The orange line gives the average contribution
of qubits $c$ and $d$ for a $4$-qubit system.}
\label{state}
\end{figure}

%\begin{figure}
%\includegraphics[width=8.5cm]{performance-comp-3q4q.png}
%\caption{Rate of success of the optimization as a function of the gate error in (a) $3$-qubit and (b) $4$-qubit systems. In (a) we show the performance of sequences with $N_p=2, 4, 8$ pulses. Solid line represents optimal protocols with all $\sigma_j \geq 0.1$;  squares, protocols with $\sigma_{j>1} \geq 0.3$; dotted lines, protocols with all $\sigma_j \ge 0.3$. In (b) the pulse sequences have $N_p = 3, 5, 7$ pulses. Solid lines represent optimal protocols with $\sigma_{j>1} \geq 0.1$, squares, protocols with $\sigma_{j>2} \geq 0.3$; dotted lines, protocols with $\sigma_j \geq 0.1$.}
%\label{persigma4}
%\end{figure}

The constraints on the proximity 
of the qubits, codified in $\sigma$, may have
relevant implications on the geometry of the arrays of atoms.  %the atoms. 
To estimate the effect of these constraints,
we perform an asymmetric optimization, where we define different
values of $\sigma_j$ (distinguished by the subindex),
%we impose \tcr{various constraints} %different constraints
%\tcr{by} defining different values of $\sigma_j$. We use 
where $\sigma_1$ is
the restriction on the minimum allowed geometrical factor of
every pulse on every qubit, $|e_{jk}| \geq \sigma_1$,
$\sigma_2$ is the second allowed minimum value, and
$\sigma_j$ refers to the $j$th minima.
%whereas $\sigma_j$, refers to the $j$th minima. 
%and we call this procedure the asymmetric-constraint optimization.
%
If all minima are the same, each qubit is treated identically,
%all qubits are treated equally,
so on average, we expect the distribution of the geometrical
factors of the optimal protocols to be similar for all qubits,
a feature characteristic of equally separated qubits (e.g. an 
equilateral triangle or a tetrahedron for $3$ and $4$ qubit systems).
On the other hand, if we chose $\sigma_1$ to be much smaller
than other $\sigma_{j>1}$, we allow for the distance
between two pairs of qubits to be larger, as in asymmetric atomic arrangements
(or isosceles triangles and linear arrays in $3$-qubit systems).

While the asymmetric optimization allows for some qubits to be used
weakly (or not at all) in optimal protocols, the qubits with the smallest
$\sigma_j$ are not predetermined,
and moreover, different pulses can use different qubits.
On the other hand, we can impose values of $\sigma$ for specific qubits,
leaving other geometrical factors unconstrained, so that 
those geometrical factors can take any value (including zero).
%in which qubits $\sigma$ \tcr{are} imposed
%and in which the geometrical factors  can take any value (including zero).
We use $\sigma_{ab}$ to refer to constraints imposed only on qubits $a$ and $b$.

\begin{figure}
\hspace*{-0.5cm}
\includegraphics[width=7.5cm]{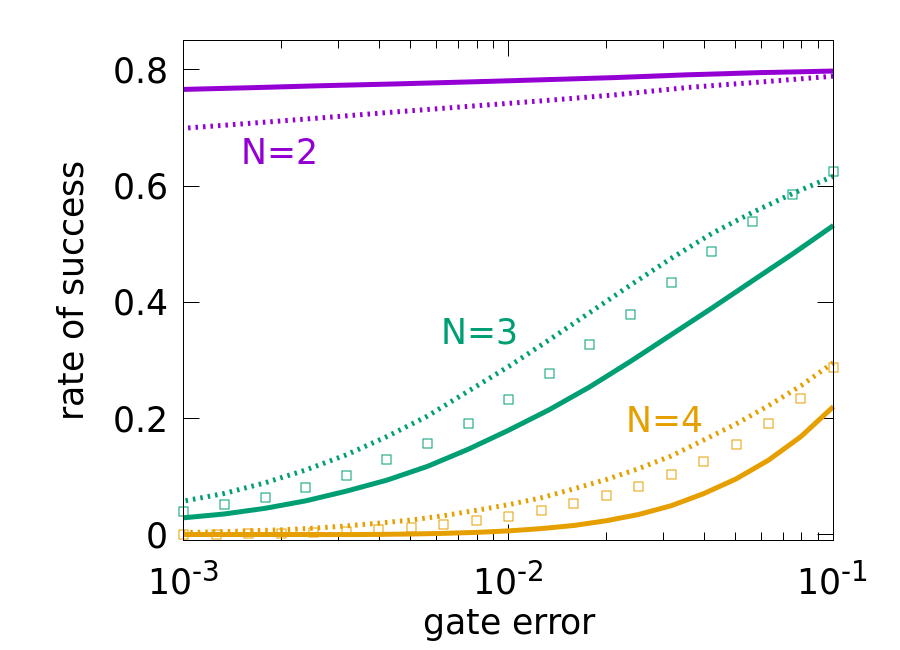}
\caption{Rate of success of the optimization for the ${\cal P}_{ab}$ gate using $6$-pulse sequences in systems with different numbers of qubits. In solid lines, we show the result when $\sigma = 0.1$. We use squares for results with $\sigma_{ab} = 0.3$ where the constrained are imposed on qubits $a$ and $b$ and the field can take any value on the remaining qubits. Dashed lines show the results when $\sigma_1 = 0$ and $\sigma_2 = 0.3$, so the constraints
are imposed in two of the three (not initially determined) qubits.}
\label{persigma}
\end{figure}

Fig.\ref{persigma} shows the rate of success of the optimization of the
${\cal P}_{ab}$ gate as a function of the gate error, using $6$-pulse sequences in systems 
with different number of qubits. 
One always observes a small decay in the rate as $\sigma$ increases 
or when more restrictive constraints are imposed,
%(or imposing more restrictive constraints) 
and a large decay when the number of qubits increases.
However, asymmetric constraints allow for better results even if $\sigma$ is
larger in a subset of qubits.
This can be inferred from Fig.\ref{persigma}
by comparing the solid lines ($\sigma =  0.1$) with the dashed
lines ($\sigma_{j>1} = 0.3$  for $3$-qubit systems, and  $\sigma_{j>2} = 0.3$
for $4$-qubit systems), where the geometrical factors on one or two qubits 
are left unconstrained, so they can be zero. 
%\tcr{que quiere decir 'where - unconstrained'}
One can easily find high-fidelity protocols enforcing the proximity of 
two of the qubits if at least the pulse does not act (or acts weakly) on the 
third (and fourth) qubits.
%Can we predetermined which qubits
%We expect qubits $a$ and $b$ to have the leading role in ${\cal P}_{ab}$. 
%Imposing constraints only on these qubits (as $\sigma_{ab} = 0.3$) 
Imposing constraints only on qubits $a$ and $b$ (as $\sigma_{ab} = 0.3$)
gives the same number of optimal protocols than the  more flexible
$\sigma_{j>1}$ asymmetric optimization at low fidelities, but less 
%optimal protocols than the more flexible  $\sigma_{j>1}$ asymmetric optimization 
as the fidelity increases.  

In $3$-qubit systems, we would then expect to obtain a large number of high-fidelity
protocols working with linear arrangements of atoms, or using geometries in isosceles triangles 
over equilateral ones.
The asymmetric optimization gives even better improvements when optimizing the
${\cal P}_{abc}$ gate. In $3$-qubit systems, all the qubits have identical roles for this gate,
but more successful optimal protocols are found when each pulse acts only (or mainly)
on two qubits at the same time.
In $4$-qubit systems we apparently do not gain much by reducing the
constraints on the parameters of the pulses over two qubits, as the results with
$\sigma_{j>2} \ge 0.3$ are not better than those with
$\sigma_{j>1} \ge 0.1$ except for very short sequences ($N_p \leq 3$).
On the other hand, forcing all pulses to act equally on 
all qubits results %redounds
in a steep decay in the performance of the protocols.
One would thus expect distorted tetrahedra or square-planar geometries or linear geometries 
to favor high fidelity protocols over tetrahedron geometries.

%However, the results show that the qubits that are not directly used in the gate ({\it e.g.}, qubit $c$ for the ${\cal P}_{ab}$ gate) are the ones that the optimal pulses use weakly. Fig.\ref{state} shows the average relative use of qubit $c$ (or of qubit $c$ and $d$ in $4$-qubit systems) in all the optimal protocols for different values of $\sigma_j$, different gates, and different number of pulses in the sequence.

To measure how the energy resources are used on average on the different qubits,
we define, for each protocol, the relative use of qubit $c$ as
\begin{equation}
    d_c = N \left( \frac{1}{N_p} \sum_k^{N_p} e_{3k}^2 \right) - 1 \ .
\end{equation}
Because ${\bf e}_k$ are normalized, if the average of the square of the
geometrical factors is $1/N$, the pulses in the optimal protocols act on 
qubit $c$ as expected from a uniform contribution, and then $d_c = 0$.
On the other hand, if $d_c \approx -1$, we can regard qubit $c$ as independent
from the other qubits.
Finally, we average $d_c$ over all the optimal protocols with fidelity higher
than $0.99$ and the results are represented as $\langle d_c \rangle$ in Fig.\ref{state}.

As expected, we observe 
$\langle d_a \rangle \approx \langle d_b \rangle \approx 0$
 in all optimal protocols.
In the ${\cal P}_{ab}$ gate, qubit $c$ (as well as qubit $d$
in $4$-qubit systems) are minimally used for short
sequences, implying that $e^2_{ck}$ barely
exceeds the imposed minimal value of the constraint,
$\sigma_1^2$ (or $\sigma_1^2 + \sigma_2^2$ in $4$-qubit
systems). However, all qubits tend to be used equally
for large sequences or when $\sigma_j \ge 0.3$.
While the constraints could allow for
a deviation $\langle d_c \rangle \sim -0.9$, this barely exceeds $-0.3$.
Optimal protocols for the ${\cal P}_{abc}$ gate use all qubits almost equally, 
except in $2$-pulse sequences.

\section{Summary and Conclusions}

We have developed models for sequences of non-overlapping pulses with control over
the spatial degrees of freedom, applied to trapped neutral atoms in ideal conditions.
We have explored in great detail the space of optimal 
protocols that implement CZ-type %CPHASE-type 
entangling gates in systems of two or more 
non-independent qubits, with high fidelity.
These qubits can potentially be much closer than in typical traps,
forming a denser quantum media that boosts the dipole blockade,
so that the gates could in principle operate in the nanosecond regime.

Studying the rate of success of the algorithm as a function of the gate error 
for different pulse sequences under different constraints, we have evaluated the 
impact of the proximity of the atoms and, indirectly, the role of the geometry 
of the arrays in $3$ and $4$-qubit systems.
To characterize the optimal protocols up to $5$-pulse sequences, we have proposed a mechanism 
analysis based on pathways that connect the initial computational state of the qubit 
with the final state, in terms of 0-, 1-, d-, and 2-loops.
We have approximately ranked the solutions in terms of
pure mechanisms, or their combinations,
characterizing each protocol by a point in a cube.

Even slight changes in the gate have a strong impact on the preferred mechanisms,
to the point that the set of optimal mechanisms mostly used  %2-qubit and 3-qubit
to implement the ${\cal  P}_{ab}$ and ${\cal P}_{abc}$ gates
barely overlap in short sequences. 
%\tcr{extranyo "Even - sentences"}
The minimal energy requirements  and the relative use of the qubits is also very different.
% *****  TALK ABOUT MINIMAL AREAS IN COMPARISON WITH THEORETICAL  RESULTS
%           FOR INDEPENDENT  QUBITS!
In the entangling $2$-qubit gate, the pulses have less impact % act less 
on the additional
qubits in the set-up, while they act equally on all qubits for the $3$-qubit gate.
However, these differences drop when the number of pulses in the sequence,
and hence the number of optimization parameters, increases.
Measuring the energy by the accumulated pulse area of all the sequence,
the first gate requires a minimum of $4\pi$ to operate with independent qubits, while
the latter needs at least $6\pi$. However, when the pulses act on two or more
qubits at the same time, the minimal total area can be substantially reduced
in gates that operate at low fidelities. This is especially so in the
case of ${\cal P}_{abc}$, when the area is distributed over $5$ or more pulses.

Asymmetric optimizations show that it is easier to find optimal
protocols with different characteristic distances
(different $\sigma_j$).
Hence, slightly asymmetric atomic arrangements in isosceles triangles 
(or linear configurations) and distorted tetrahedra (or squares)
allow many more high-fidelity protocols than the more symmetric
structures.

We have shown that it is always possible to find high-fidelity protocols
even with short sequences in systems with more than $2$-qubits. 
The rate of success is smaller when the
number of pulses diminishes or the qubits are closer, 
with worse results when we impose large fields everywhere. 
Given that the number of parameters that control the system raises with 
the number of qubits and pulses as $N \times N_p$, % $N_p$, while the constraints imposed by 
while the gate implementation requires an exponential increase
of constraints, $2^N$, it is somehow
surprising that one can obtain relatively high fidelity solutions
($\epsilon < 0.99$) using $N_p \sim N +1$.
We have also found that by increasing the allowed pulse areas
one can find solutions even for highly interacting qubits.

The general findings in this work allow us to speculate that one could work with 
denser arrays of qubits boosting the dipole blockade, %and 
accelerating the operating time of the gates by almost two orders of magnitude,
and hence nearly reaching the nanosecond time-scales.
Fast gates are inherently more robust to decoherent effects.
Preliminary studies show that the gates are also relatively robust to the
thermal motion of the atoms and %as well as 
fluctuations in the pulse intensities.
However, further studies are needed to assess %nonlinear effects 
the effect of nonlinear effects 
in the Hamiltonian, not included in our models, as well
as the practical limitations in moving the atoms or using 
complex light structures in the spatial-domain, instead of the
time-domain.

\section*{Acknowledgements}
This research was supported by the Quantum Computing Technology Development Program (NRF-2020M3E4A1079793). IRS thanks the BK21 program (Global Visiting Fellow) for the stay during which this project started and the support from MINECO PID2021-122796NB-I00. SS acknowledges support from the Center  for Electron Transfer funded by the Korea government(MSIT)(NRF-2021R1A5A1030054)

\bibliography{maintext.bib} %You need to replace "rsc" on this line with the name of your .bib file

\end{document}